\documentclass[aps,prb,twocolumn,superscriptaddress,floatfix]{revtex4}
\usepackage{epsfig,amsmath,amssymb,color}
\usepackage{graphicx}

\begin{document}

\title{Low temperature specific heat of 1D multicomponent systems at commensurate- incommensurate phase transition point}

\author{T. Vekua}
\affiliation{Institut f\"{u}r Theoretische Physik, Leibniz Universit\"at Hannover\\ 
Appelstra\ss e 2, 30167 Hannover, Germany}

\date{\today}
    
\begin{abstract}

Low temperature dependence of specific heat of one- dimensional multicomponent
systems at the commensurate- incommensurate phase transition point is
studied. It is found that for canonical systems, with a fixed total number of
particles, low temperature specific heat linearly depends on temperature with
a diverging prefactor. 
\end{abstract}

\maketitle

Second order phase transitions in one- dimensional quantum or two- dimensional
statistical systems are often classified by fixed point Hamiltonians enjoying
conformal invariance\cite{BPZ}. In such situations, low temperature
thermodynamic behavior is fixed solely by the central
charge\cite{DiFranchesco}, $c$, the quantity that roughly measures numbers of
degrees of freedom, and velocity of the linear excitation, $v$ - `speed of
light'. Low temperature specific heat combines these two quantities and has
the universal form\cite{Cardy} $C(T)=\pi c T/3 v$, depending linearly on temperature.
The notable exception from the above picture is provided by the second order
phase transition driven by chemical potential coupled to the conserved
quantity\cite{Sachdev0}, so called commensurate- incommensurate (C-IC) phase
transition\cite{Japaridze,Pokrovski}. At the C-IC phase transition point
system, in contrast to conformal invariance, obeys zero scale- factor universality\cite{Sachdev} with dynamical critical exponent $z=2$. In particular for the specific heat it implies square root temperature dependence for $T\to 0$. 

Describing the C-IC phase transitions in multicomponent systems (systems of
fermions, bosons or spin $S$ chains) one can not rely on mode- decoupling
approximation\cite{Vekua1}. In particular, as a result of spin- charge non-
separation, magnetic susceptibility of spin gapped systems, at the edge point
of magnetization plateau, stays finite instead of diverging  unless special microscopic symmetries
are present (e.g. particle-hole symmetry at half filling for attracting Fermi
Hubbard model). This kind of behavior has also been observed for a number of
integrable models such as the two component Fermi Hubbard model both on lattice\cite{Woynarovich86,FrahmVekua} as well as in continuum\cite{Orso}, or spin $S$ generalization of the integrable $t-J$ chain doped with $S-1/2$ carriers\cite{Frahm}.  
Similar behavior holds for generic systems of multicomponent Fermi or Bose
gases, with gapped spin and gapless charge excitation\cite{Vekua1,Vekua2}. Notable
example of bosonic systems, in this respect, is provided by antiferromagnetically interacting spin 1 Bose gas\cite{Vekua2,Guan}. 

 Zero temperature magnetization increases linearly with the field at the edge 
 of the magnetization plateau, due to the finite value of magnetic susceptibility.
This kind of a behavior is characteristic to conformally invariant systems. The natural question
 arises whether some kind of conformally invariant theory can be applied to
 describe the edge points of magnetization plateaus. For this reason we study
 temperature dependence of specific heat in multicomponent systems at C-IC
 phase transition point where modes do not decouple. We find that if total number of particles is fixed then square root temperature dependence, which holds if one works with fixed chemical potential and also is characteristic to mode- decoupled systems undergoing C-IC transition (for both cases with fixed chemical potential or fixed total number of particles, e.g. attractive Hubbard model at half filling at the onset of magnetization), is modified by linear law, however the prefactor diverges as $\ln^2{T}$ for $T\to0$. To our knowledge no previous studies on finite temperature thermodynamic behavior at C-IC phase transition point have been reported for multicomponent systems where modes do not decouple.

For concrete calculations we will consider a system of two component
attractive fermions and place it in an external magnetic field equal to the
value of the spin gap. However our main claim on the disappearance of the square
root temperature dependence of the specific heat (at C-IC phase transition
point) will hold true for generic situation, where gapped mode (spin), undergoing non- relativistic softening at the critical point, couples to linearly dispersing excitations (charge). This statement will apply to all situations where magnetic susceptibility stays finite at the edge points of magnetization plateaus in zero temperature limit\cite{Vekua1}. 

To simplify things drastically we will consider the dilute limit of the
system of attractively interacting two component fermions. Since the dilute limit falls under strong coupling, the ground state
(for $T=0$) is made of bound pairs and the low temperature thermodynamic
properties can be modeled by the mixture of noninteracting pairs and thermally
created un-compensated spin- up particles (we choose finite temperature
magnetization to be positive for nonzero magnetic field). Spin- down particles
will also be created thermally, but since they have spectral gap (which is
large in strong coupling) their density will be exponentially suppressed at
low temperatures and thus they will be ignored in the following. For
integrable Fermi Hubbard model (both on lattice\cite{Essler} and in continuum)
the thermodynamic Bethe ansatz method can be applied\cite{Takahashibook}, and in particular recently it was shown that in dilute limit this method simplifies considerably\cite{Zhao}. However, for finding specific heat at the critical point when magnetization sets in we need not use integrability so that our reasoning will stay both simple and general.

Phenomenological Hamiltonian describing the low energy properties of the two component attractive fermions in strong coupling at the C-IC critical point reads:
\begin{equation}
\label{PhenHam}
 H=\sum (E_p(k)-E_F)a^{\dagger}_{k} a_{k} +\sum E_{\uparrow}(k)c^{\dagger}_{k} c_{k},
\end{equation}
where $a^{\dagger}(c^{\dagger})$ and $a(c)$ are pair (up- spin fermion) creation and annihilation operators. 
For calculating thermodynamic properties it does not matter whether pairs are modeled by hard core bosons (which they are) or fermions\cite{Takahashibook}.

\begin{figure}
\begin{center}
  \includegraphics[scale=0.6]{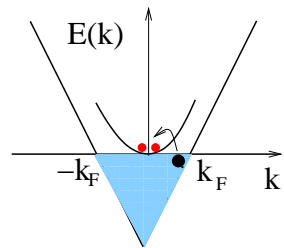}
\end{center}
\caption{simplified dispersions with purely quadratic and linear behavior adopted in this paper. If total number of particles is conserved creation of each pair of up- spin fermions (smaller bullets) leaves a hole in the Fermi sea of bound pairs (larger bullet). }
\label{fig:dispersions}
\end{figure}

At the edge of the zero magnetization plateau dispersion of up- spin particles
is approximated by a purely quadratic expression:
\begin{equation}
 E_{\uparrow}(k)=\sqrt{v_{\uparrow}^2k^2+\Delta^2}-\Delta\simeq v_{\uparrow}^2 k^2/2\Delta.
\end{equation}
 Here $\Delta$ is spin gap opened due to strong attraction between spin- up
 and spin- down fermions, and $v_\uparrow$ is the velocity of gapped
 excitations which, as we will see later, will not appear in the leading temperature dependence of
 specific heat. Since low
 temperature thermodynamics is determined by low lying modes (states near
 Fermi wave- vector) for the dispersion relation of pairs we can use a simplified strictly linear expression: 
\begin{equation} 
E_p(k)-E_F=v_p(|k|-k_F).
\end{equation}
 Fermi wave vector $k_F$ is related to the linear density of pairs $n_p$ by
 the usual 1D relation $k_F=n_p \pi$. In the following $n_p$ will stand for the density of pairs strictly at zero temperature: $n_p=n_p(T=0)$.
Dispersions $E_{\uparrow}(k)$ and $E_p(k)-E_F$ considered in our simplified model are plotted on Fig.(\ref{fig:dispersions}).
At zero temperature the ground state is occupied solely by pairs, and density of un-compensated up- spin electrons is zero. 
Although one might argue that Hamiltonian (\ref{PhenHam}) lacks microscopic
 derivation, it is expected to exactly reproduce leading low temperature dependence of the specific heat at C-IC phase transition point in dilute limit, the same way as it yields exact expression for magnetic susceptibility\cite{Vekua1}.
The thermodynamic potential of the grand canonical system can be obtained straightforwardly:
\begin{equation*}
\frac{\tilde\Omega}{L}=-\frac{T}{2\pi} \int dk [ \ln(1+e^{ -\frac{v_p(|k|-k_F) }{T}  }) +\ln(1+ e^{-\frac{ v_{\uparrow}^2 k^2/2\Delta}{T}} )].
\end{equation*}
For specific heat one obtains easily that the leading
temperature dependence is linear from pairs (usual conformal
contribution\cite{Cardy} with central charge $c=1$ and velocity of pair motion
$v_p$) and square root (which dominates over the linear one in the low
temperature limit), from the up- spin electrons, due to the quadratic low
energy dispersion they enjoy. This kind of behavior is also realized for
canonical systems at special fillings, e.g. the attractive Fermi Hubbard model
at half filling, where spin and charge modes decouple. At half filling charge sector of Hubbard model is
described by conformally invariant $SU_1(2)$  Wess- Zumino model\cite{Essler}, while the
spin sector, when magnetic field equals to singlet- triplet spin gap, is
characterized by quadratic low energy behavior responsible for the leading,
square root, temperature dependence of specific heat. 
In the grand canonical systems at finite temperatures, the thermal fluctuations
create a finite density of up-spin electrons given by the following expression: 
\begin{equation}
\label{numberup}
 \tilde n_{\uparrow}(T)=-\zeta(1/2) \sqrt{2\pi \Delta} (\sqrt{2}-1) \sqrt{T}/2\pi v_{\uparrow},
\end{equation}
where $\zeta(1/2)$ is the value of the Riemann $\zeta$ function at $0.5$.
 The number of pairs also gets modified by temperature fluctuations accordingly:
\begin{eqnarray}
\label{numberpair}
\tilde n_p(T)&=&n_p+\frac{T}{\pi v_p}\left(e^{-v_pk_F/T} +O( e^{-2v_pk_F/T} )\right).
\end{eqnarray}
Thus, due to finite ground state population of pairs, thermal fluctuation of density of pairs is exponentially suppressed, whereas for up- spin electrons, due to Van Hove singularity in their density of states, square root dependence holds.
Already from the above picture it is clear that working at a fixed total number of particles can modify the leading square root behavior of specific heat at C-IC point, since \textit{thermal occupation of up- spin band will be constrained by the reluctance of depletion of the band of bound pairs}.
Since each broken pair produces two up- spin electrons, as depicted on Fig.(\ref{fig:dispersions}), the constraint that fixes the total number of particles reads:
\begin{equation}
\label{con}
n_{\uparrow}(T)+2n_p(T)=2n_p.
\end{equation}

To implement the constraint we introduce the temperature dependent chemical potential, which we determine from the equation:
\begin{equation}
\int\!\!\! \frac{dk/2\pi}{ e^{\frac{ v_{\uparrow}^2 k^2/2\Delta-\mu(T) }{T}} +1}+\int\!\!\!\frac{dk/\pi}{ e^{\frac{ v_p(|k|-k_F)-2\mu(T) }{T}} +1}=2n_p.
\end{equation}
The consistent solution of the above equation requires that $ \mu(T)/T\to -
\infty$, when $T\to 0$, so that we obtain: 
\begin{equation}
 {\mu(T)}\simeq -\alpha \sqrt{T}e^{\mu(T)/T},
\end{equation}
with positive coefficient $\alpha=\frac{v_p\sqrt{2\Delta \pi}}{8 v_{\uparrow}}$.
In the zero temperature limit we obtain the following solution for the chemical potential:
\begin{equation}
\label{chemicalpotential}
 \mu(T)\simeq -T(\ln{\frac{\alpha}{\sqrt{T}}} -\ln \ln{\frac{\alpha}{\sqrt{T}}}+ \frac{\ln \ln{\frac{\alpha}{\sqrt{T}}}}{\ln{\frac{\alpha}{\sqrt{T}}}}).
\end{equation}

Now we will use the obtained chemical potential Eq.(\ref{chemicalpotential}) to calculate the grand canonical potential:
\begin{eqnarray}
\frac{\Omega}{L}=&-&\frac{T}{2\pi} \int dk \ln(1+e^{ -\frac{v_p(|k|-k_F)-2\mu(T) }{T}  })\nonumber\\
& -& \frac{T}{2\pi}\int dk \ln(1+ e^{-\frac{ v_{\uparrow}^2 k^2/2\Delta-\mu(T) }{T}} )\nonumber\\
=&-&\frac{T^2}{2\pi v_p} \left(  \frac{(v_pk_F+2\mu(T))^2}{T^2}+\frac{\pi^2}{3}+\cdots\right)\nonumber\\
&-&\frac{T^{3/2}}{2\pi v_{\uparrow}} \sqrt{2\pi \Delta} e^{\mu(T)/T}+\cdots
\end{eqnarray}
 Dots stand for sub-leading terms at low temperatures.
Grand canonical potential is related to energy, $E$, and entropy $S$, by the following thermodynamic relation:
\begin{equation}
\label{Therm}
\Omega=E-TS-\mu(T) N_{\uparrow}(T)-(v_pk_F+2\mu(T))N_p(T).
\end{equation}
For the leading temperature dependence of energy we get the following expression:
\begin{equation}
 \frac{E}{L}=\frac{(v_p k_F+2\mu(T))^2}{2\pi v_p}+\frac{\pi T^2}{6 v_p}-\frac{2T\mu(T)}{\pi v_p}+\cdots
\end{equation}
 For thermally depleted density of pairs we obtain:
\begin{equation}
 n_p(T)=\frac{v_pk_F+2\mu(T)}{\pi v_p},
\end{equation}
and the density of thermally created up- spin electrons is given by the following expression:
\begin{equation}
n_{\uparrow}(T)=2n_p-2n_p(T).
\end{equation}
Note, due to constraint of keeping total number of particles fixed, the square root dependence Eq.(\ref{numberup}) is modified in the following way: 
\begin{equation}
n_{\uparrow}(T)\simeq -\frac{2T\ln{(T/\Delta)}}{\pi v_p},\,\,\,\,\,\, T/\Delta \ll 1.
\end{equation}

Using thermodynamic relation Eq.(\ref{Therm}) we obtain for the leading temperature dependence of entropy the following expression:
\begin{equation}
s(T)=S(T)/L=\frac{4\mu^2(T)}{T\pi v_p}-\frac{6\mu (T)}{\pi v_p}+\frac{\pi T}{3 v_p}+\cdots
\end{equation}

Finally substituting chemical potential from Eq.({\ref{chemicalpotential}}) for the leading temperature dependence of specific heat we get:
\begin{equation}
\label{mainresult}
 C(T)=T\frac{\partial s(T)}{\partial T}= \frac{T\ln^2{(T/\Delta)}}{\pi v_p}+\cdots
\end{equation}
 in the limit $T\to 0$.
This is our main result: specific heat depends linearly on temperature, albeit
with a diverging prefactor. This is kind of an intermediate between the square
root and linear dependences. Recent studies reported similar linear behavior
of the specific heat with diverging prefactor for critical two- subband
quantum wire\cite{Rosch} when the second band started to fill. However in our case we can not interpret this as a signal of restoration of Lorentz or conformal invariance.

At this point we would like to comment on the differences between working at fixed chemical potential and fixed total number of particles  for attractive Hubbard model away of half filling at the onset of magnetization (C-IC transition point).
For the case of fixed total number of particles magnetic susceptibility stays finite and, as we showed here, specific heat 
depends linearly on temperature (with diverging prefactor). However, for the case of fixed chemical potential magnetic susceptibility diverges and specific heat shows square root temperature dependence- characteristic to the mode decoupled theory. One may suspect that if one works with the fixed chemical potential one can use mode decoupled theory, result of linear bosonization, to describe the C-IC phase transition point. For calculating static magnetic susceptibility and specific heat this statement is true. However, using mode- decoupled theory across the C-IC transition is misleading, because away of half filling and for non-zero magnetization spin and charge modes couple independent of whether one fixes chemical potential or total number of particles. In particular spin and charge stay coupled (spin- charge mixing angle stays \textit{finite}) even in the limit of vanishing magnetization (C-IC point) for both cases with fixed total number of particles or fixed chemical potential. Within spin- charge decoupled theory one can not calculate correctly e.g. assymptotics of static single particle correlation functions\cite{Vekua1} in the limit of vanishing magnetization (C-IC point). Fixing total number of particles further enhances mode coupling and leaves its traces on magnetic susceptibility as well as specific heat (topic of this manuscript).

The presented analysis is directly applicable also to the phase transition
between partially and fully polarized states, when the down- spin
fermions band gets empty (at $h=h_{sat}$, where $h_{sat}$ is magnetic field
needed to completely polarize the system). In this case one simply has to replace constraint Eq.(\ref{con}) by: $N_{\uparrow}(T)+N_{\downarrow}(T)=const$, thus specific heat at $h=h_{sat}$ will have the similar temperature dependence as in Eq.(\ref{mainresult}).

For completeness we write out the leading low temperature dependence of
specific heat in two component attractively interacting fermions in an
external magnetic field for all values of the field away from half filling: 
\begin{equation}
\label{completespecificheat}
C(T)\sim  \left\{ \begin{array}{lr}
  T/{v_c} & h<2\Delta \,\,\,{\mathrm or}\,\,\, h>h_{sat}\\
T\ln^2 {(T/\Delta)} & h=2\Delta \,\,\,{\mathrm or}\,\,\, h=h_{sat}\\ 
T(v_c+v_s)/v_cv_s &2\Delta<h<h_{sat}
\end{array} \right .
\end{equation}
where $v_c$ and $v_s$ denote charge and spin velocities\cite{comment0}. Note that
$v_s(m)\sim m$ when $m\to 0$, and  $v_s(m)\sim n-2m$ when $m\to n/2 $, where $m$
is the magnetization density and $n$ the total density of particles.  In the region $\Delta<h<h_{sat}$ spin and charge modes are coupled,
and they do not get separated even in the limits $h\to 2\Delta_+$ and $h \to
h_{sat_-}$\cite{Vekua1,FrahmVekua}. This non- separation of
spin and charge modes explains disappearance of the square root temperature dependence of specific heat at critical points $h=2\Delta$ and $h=h_{sat}$\cite{comment}. 
Exactly at half filling, where spin- charge separation holds, $C(T)\sim \sqrt {T}$ for $h=2\Delta\,\, {\mathrm or}\,\,
h=h_{sat}$.

To summarize on simple example of two component attractive fermions placed in
 an external magnetic field equal to the spin gap we studied the thermodynamic
 properties of C-IC phase transition point in multicomponent systems. Our
 findings are generic, namely, we expect that for multicomponent systems square root temperature dependence, which holds for the fixed chemical potential, will be changed by linear dependence (with diverging prefactor) if instead total number of particles is fixed. 
In particular, this
 statement equally applies to the edge points of magnetization plateaus of 1D bosonic systems, i.e. spin 1 bosons interacting antiferromagnetically\cite{Zhou}. It also holds for the edge points of mid magnetization plateaus of multicomponent systems, i.e. in doped dimerized Hubbard chains\cite{Cabra}.
 However, currently we can not determine if the $\ln^2T$ prefactor in the
 expression of specific heat Eq.(\ref{mainresult}) is universal. To answer
 this question, it will be useful to study e.g. the same system of two
 component attractive fermions in the weak- coupling limit. One can make use of integrability of Fermi Hubbard model there. Especially intriguing will be to look close to half filling to follow how does the square root dependence of specific heat cross over to linear behavior with diverging prefactor as a function of doping.

Our findings may be relevant for real quasi one-dimensional spin gapped
materials, where one can study experimentally specific heat as a function of
applied magnetic field and doping. Square root dependence, which is expected
at half filling, should be modified by linear with logarithmic prefactor, away
from half filling, at critical magnetic field strength equal to the spin
gap. One can as well measure the leading temperature dependence of specific heat at saturation 
magnetization in a generic 1D electron system away of half filling and compare the results to Eq.(\ref{mainresult}).

\section{Acknowledgments}

The author is supported by cluster of excellence QUEST: center for quantum engineering and space-time research.


\begin{thebibliography}{10}
\bibitem{comment0} More rigorously, in the region $\Delta<h<h_{sat}$, $v_{c,s}$ are velocities of the modes which are combinations of spin and charge degrees of freedom that diagonalize the two component Luttinger Liquid Hamiltonian and are denoted by $v_{\pm}$ in \cite{Vekua1}.
\bibitem{comment} We remind that in 1D systems critical points exist strictly
  at zero temperature. At a finite temperature magnetization sets in for an
  arbitrary nonzero magnetic field and the saturation field is shifted to infinity.
\end{thebibliography}
\end{document}